\documentclass[aps,floatfix,prd,twocolumn]{revtex4}
\usepackage{graphicx}
\usepackage{dcolumn}
\usepackage{bm}
\usepackage{amsmath}
\usepackage{color}

\def\*{^{(*)}}

\def\q{\mathbf q}
\def\vs{\boldsymbol \sigma}
\def\vt{\boldsymbol \tau}

\def\vq{\mathbf q}

\def\Lc{\Lambda_c}
\def\LL{\Lambda_{c}(2595)}
\def\Ll{\Lambda_{c}(2595)}
\def\S{\Sigma_c}
\def\Lt{\tilde\Lambda}

\def\>{\big>}
\def\<{\big<}
\def\|{\big\vert}
\def\£{\big\Vert}

\def\*{^{(*)}}

\def\bit{\begin{itemize}}
	\def\eit{\end{itemize}}

\def\d0{D\O{}}

\def\D{\bar D}

\def\*{^{(*)}}
\def\etal{\textit{et~al.}}
\def\icc{$\S\D^*-\Ll\D$}
\def\ecc{$\S\D^*-\S\D^*$}
\def\SD{\S\D^*}
\def\LD{\Ll\D}

\begin{document}

\title{Molecular Interpretation of the $P_c(4440)$ and $P_c(4457)$ States}
\author{T.J. Burns}
\affiliation{Department of Physics, Swansea University, Singleton Park, Swansea, SA2 8PP, UK.}
\author{E.S. Swanson}
\affiliation{Department of Physics and Astronomy,
University of Pittsburgh,
Pittsburgh, PA 15260,
USA.}

\begin{abstract}
A molecular model of the $P_c(4457)$ and $P_c(4440)$ LHCb states is proposed. The model relies on channels coupled by long range pion-exchange dynamics with features that depend crucially on the novel addition of the $\Lambda_c(2595)\bar D$ channel.  A striking prediction of the model is the unusual combination of quantum numbers $J^P(4457) = 1/2^+$ and $J^P(4440) = 3/2^-$. Unlike in other models, a simultaneous description of both states is achieved without introducing additional short-range interactions. The model also gives a natural explanation for the relative widths of the states. We show that the usual molecular scenarios cannot explain the  production rate of $P_c$ states in $\Lambda_b$ decays, and that this can be resolved by including $\Lambda_c(2595)\bar D$ and related channels. Experimental tests and other states are discussed in the conclusions.
\end{abstract}

\maketitle

\section{Introduction}

The apparently exotic nature of the $P_c(4380)$ and $P_c(4450)$ baryons discovered at LHCb in 2015 \cite{Aaij:2015tga,Aaij:2016phn} has provoked intense interest in the hadron physics community. Recently, LHCb reported on the analysis of $\Lambda_b\to J/\psi\, p \,K^-$ with an order of magnitude more data~\cite{Aaij:2019vzc}, and the situation has now become even more interesting. What was previously the $P_c(4450)$ is actually two distinct states, $P_c(4440)$ and $P_c(4457)$, and there is an additional new state, $P_c(4312)$. (Due to its width, $P_c(4380)$ was not visible in the recent analysis, which is sensitive only to narrow states.)

All of the states are observed in $J/\psi\, p$, which suggests that their wavefunctions are composed, in some way, of the combination of quark flavours $uudc\bar c$. Models differ in  whether the relevant degrees of freedom are the five constituent quarks, effective diquark combinations such as $ud$ and $dc$, or meson-baryon combinations, either with closed flavour $(uud)(c\bar c)$, or open-flavour $(udc)(u\bar c)/(uuc)(d\bar c)$.

\begin{table}
\begin{tabular}{llll}
\hline\noalign{\smallskip}
State&Mass /MeV & Thresholds /MeV &\\\noalign{\smallskip}
\hline\hline\noalign{\smallskip}
$P_c(4312)$&$4311.9\pm0.7^{+6.8}_{-0.6}$&$4317.7\pm 0.45$& $(\S^+\D^0)$\\\noalign{\smallskip}
\hline\noalign{\smallskip}
$P_c(4380)$&$4380\pm 8\pm 29$&$4382.3\pm 2.4$&$(\S^{*+}\D^0)$\\\noalign{\smallskip}
\hline\noalign{\smallskip}
$P_c(4440)$&$4440.3\pm1.3^{+4.1}_{-4.7}$&$4459.75\pm0.45$&$(\S^+\D^{*0})$\\\noalign{\smallskip}
$P_c(4457)$&$4457.3\pm 0.6^{+4.1}_{-1.7}$&
$4457.08\pm0.33$&$(\Lc(2595)\D^0)$\\\noalign{\smallskip}
\hline\hline\noalign{\smallskip}
\end{tabular}
\caption{The masses of $P_c$ states and nearby thresholds.}
\label{tab:thresholds}
\end{table}

Experimental data suggests the last of these scenarios is favoured. As shown in Table \ref{tab:thresholds}, the masses of all four of the states are in close proximity to two-body thresholds for open-flavoured hadron pairs, which is a strong indication that such pairs are the relevant degrees of freedom. In this picture the states could be hadronic resonances of molecular nature (which we advocate in this paper), or threshold effects such as cusps or triangle singularities.

The simplest molecular scenario, where binding arises due to pion exchange in the elastic channel (no coupled channel effects), is inadequate in the sense that it leads to molecular states only with $\S\D^*$ and $\S^*\D^*$ constituents, thus accounting for only one of the four thresholds ($\S\D^*$) identified in Table~\ref{tab:thresholds}. Another problem, which has not been discussed in the literature, is that in $\Lambda_b$ decays (where all of the $P_c$ states are observed) the diagrams producing the hadron pairs of three of the four nearby thresholds ($\S\D$, $\S^*\D$, and $\S\D^*$) are strongly suppressed, whereas other pairs (whose thresholds are not so near to $P_c$ states) are produced more strongly. Finally, the LHCb replacement of $P_c(4450)$ with two distinct states, $P_c(4440)$ and $P_c(4457)$, is a challenge for  models without coupled channels, as these predict only one state in this mass region.  

Incorporating coupled-channel effects can potentially resolve all of these problems. It implies that there is no longer an automatic restriction on which open-flavoured channels can support molecular states, so that in principle, all four of the thresholds identified in Table~\ref{tab:thresholds} can be relevant. As we shall show, it also gives a simple solution to the issue of the production mechanism for all four $P_c$ states.

Our main focus in this paper is on the remaining  problem, namely the need to account for two states  near $\S\D^*$ threshold. The pion-exchange model with only $\S\D^*$ degrees of freedom predicts the existence of only one state, with $3/2^-$ quantum numbers. (The potential in the $1/2^-$ channel is repulsive and does not support a bound state.) The need to account for another state suggests that some extra degrees of freedom, previously neglected, need to be included in the model.

Our proposal is that the missing ingredient, whose inclusion can resolve this problem, is the $\Ll\D$ channel. It is based on the idea, initially proposed by one of us, to interpret $P_c(4450)$ as a coupled-channel \icc~ state, similar in some respects to the $X(3872)$ as a $D\D^*-D^*\D$ state \cite{Burns:2015dwa}. The channels $\S\D^*$ and $\Ll\D$ are coupled by one-pion exchange, and due to the remarkable proximity of the two thresholds (see Table~\ref{tab:thresholds}), this coupling should logically be included in any molecular scenario. 

The recent results from LHCb give extra impetus to this idea, not only because the existence of two states suggests the need for extra degrees of freedom, but also because, whereas the old $P_c(4450)$ had mass around 7~MeV below $\Ll\D$ threshold, the new state $P_c(4457)$ coincides exactly with $\Ll\D$ threshold (see Table~\ref{tab:thresholds}), which is a strong indication of $\Ll\D$ degrees of freedom in its wavefunction.

As well as accounting for both $P_c(4440)$ and $P_c(4457)$, our model makes an unambiguous prediction for their quantum numbers, which  are $3/2^-$ and $1/2^+$, respectively. Currently, there is no experimental information on $J^P$ for the $P_c$ states; the previous amplitude analysis gave preferred assignments for $P_c(4380)$ and $P_c(4450)$, but these  are now considered obsolete, as they were based on a two-state fit to data. Thus it is up to models to make predictions for $J^P$, which can ultimately be tested in future experimental analyses. Our prediction of $1/2^+$ quantum numbers of $P_c(4457)$ is particularly novel, as almost all competing models  assign this state to $1/2^-$ or $3/2^-$. The opposite parity in our model arises because $\Ll$, as an orbitally-excited state, has opposite parity to the usual molecular constituents.

Our model is also consistent with experimental measurements of the production and decay properties of the $P_c$ states. Analysis of photo-production cross-sections implies the states must decay prominently to channels other than $J/\psi\, p$ \cite{Wang:2015jsa,Ali:2019lzf}; in the molecular scenario the missing channels are open-flavoured hadron pairs and their partial widths are calculable in the same formalism as is used for the molecular binding. We find, in particular, that simple selection rules explain why the  $P_c(4457)$ and $P_c(4440)$ are comparatively narrow and broad, respectively. We can also account for the comparable experimental values for the product of the production and decay branching fractions of the two states.

The plan of the paper is as follows. In Section~\ref{sec:model}, we introduce our model, and show that it gives a striking and unusual prediction for the quantum numbers of $P_c(4440)$ and $P_c(4457)$. We demonstrate in Section~\ref{sec:binding} that pion exchange in the coupled channel \icc~system  generates states consistent with $P_c(4440)$ and $P_c(4457)$. In Section~\ref{sec:decays} we show that our model naturally accounts for the relative widths of the two states, and in Section~\ref{sec:isospin} we argue that, unlike other molecular scenarios, in our model isospin mixing is negligible. In Section~\ref{sec:production} we show that the production of $\S\*\D\*$ states in $\Lambda_b$ decays is suppressed, which is a challenge for most models of $P_c$ states, and a possible indication of the importance of coupled-channel effects: we argue that including $\Ll\D$ and related channels channel can resolve this problem. We conclude in Section~\ref{sec:conclusion}.

\section{The model}
\label{sec:model}

Pion exchange in the elastic channel cannot explain all four of the $P_c$ states, since it can generate $\S\D^*$ and $\S^*\D^*$ states, but not $\S\D$ or $\S^*\D$ states. A natural extension of the model, which is anyway required for self-consistency, is the inclusion of the one-pion exchange coupling between different hadron constituents.  Just as it couples elastic channels (such as $\S\D^*\to\S\D^*$), one-pion exchange also couples inelastic channels (such as  $\Lc\D^*\to \S\D$). When all such channels are included in the calculation, molecular states can potentially appear at the thresholds for any open-flavoured pair, including all of those identified in Table~\ref{tab:thresholds}. 

Most existing work in this area considers coupling among $\S\*\D\*$ channels \cite{Shimizu:2016rrd,Shimizu:2018ran,Shimizu:2019ptd,Shimizu:2019jfy,He:2019ify}, or both $\S\*\D\*$ and  $\Lc\D\*$ channels \cite{Wu:2012md,Yamaguchi:2016ote,Yamaguchi:2017zmn,Shimizu:2017xrg,Chen:2019asm,Huang:2015uda,Yang:2015bmv,Huang:2019jlf,Huang:2018wed}, but the $\LL\D$ channel has hardly been discussed.  There is some variation among the predictions of the different papers, but several identify states near the $\S\D$, $\S^*\D$, and $\S\D^*$ thresholds which are matched with $P_c(4312)$, $P_c(4380)$ and $P_c(4440/4457)$, respectively. (For $P_c(4312)$, a data-driven analysis \cite{Fernandez-Ramirez:2019koa} indicates a virtual state nature.)

A  common feature of these calculations, which is consistent with naive expectations, is that molecular states are most likely in $J^P$ channels where the dominant channel is in S-wave. For the $P_c(4440)$ and $P_c(4457)$, assuming their dominant components are $\S\D^*$, the allowed quantum numbers are therefore $1/2^-$ or $3/2^-$, and indeed this is what is found (or assumed) in most molecular models for these states. But as we argue later in this paper, these assignments are not tenable with only $\S\D^*$ constituents bound by pion exchange alone. We find, consistent with other authors, that a $3/2^-$ naturally arises, but not a $1/2^-$ state \cite{Karliner:2015ina,Chen:2015loa,Chen:2016heh,Takeuchi:2016ejt,Eides:2017xnt}.

The $\Ll\D$ channel is conspicuously absent from most of the discussion on coupled-channel systems, which is surprising considering that its threshold coincides exactly with the $P_c(4457)$ mass. This channel does not experience elastic one-pion exchange, but it couples inelastically to other channels, the most important of which will be $\S\D^*$, on account of the proximity of the thresholds (see Table~\ref{tab:thresholds}). 

The idea of a molecule arising from the \icc~ coupling was first proposed in ref.~\cite{Burns:2015dwa}, using phenomenological arguments. As far as we are aware, there is only one paper in the literature which has considered this system quantitatively, which is that of Geng~\etal~\cite{Geng:2017hxc}. They derived the relevant ``vector'' potential which, with its $1/r^2$ dependence at short-distance, is in a sense intermediary between the more familiar central and tensor potentials. Their main motivation for studying the \icc~  system was the interesting observation that portions of the system could exhibit discrete scale invariance under certain conditions. (It turns out that, with physical parameters, this property is not quite realised in this system, but some other systems where it does arise were highlighted.) The scale invariance idea derives from the leading $1/r^2$ dependence of the inelastic \icc~ potential: the corresponding elastic potential \ecc does not have this property, and thus was not included in the calculations of ref.~\cite{Geng:2017hxc}.

\begin{figure*}
\includegraphics[width=0.60\textwidth]{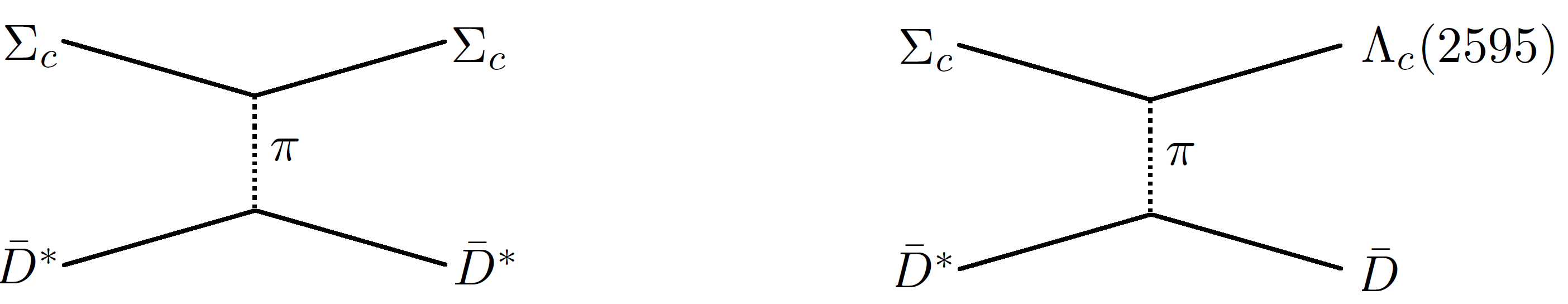}
\caption{Elastic and inelastic $t$-channel pion exchange diagrams in the \icc~system.}
\label{fig:ope}
\end{figure*}

Unlike Geng \etal, in this paper we include both the elastic and inelastic couplings, and refer to this collectively as the ``\icc~ system''~(see Fig.~\ref{fig:ope}). This makes an important difference: it is only by including both contributions that we obtain a simple and natural explanation for the existence of two states in the relevant mass region. 

Pion vertices are P-wave transitions in the usual  molecular scenarios with  $\Lc\D\*$ and $\S\*\D\*$ channels since all of the constituents are ground-state (S-wave) hadrons. But for the \icc~ coupling, $\S$ is an S-wave state, while $\Ll$ is P-wave, so that the $\S\Ll\pi$ vertex is S-wave. In a different context, Close~\etal~\cite{Close:2010wq,Close:2009ag} proposed that molecular states with both S- and P-wave constituents (hence S-wave vertices) could be particularly susceptible to binding. The constituents in the cases they considered have large width, and it was argued by others \cite{Filin:2010se} that this makes it difficult to form molecular states. This is not a problem in our case, as both $\S$ and $\Ll$ are narrow resonances, due to very little phase space in their decays~\cite{Burns:2015dwa}.

Without doing any calculations, we can see why the \icc~ system can naturally account for the existence of two molecular states in the mass region of $P_c(4440)$ and $P_c(4457)$, and additionally, we can anticipate their $J^P$ quantum numbers. We obtain an unusual quantum number assignment because of the opposite parities of the $\S$ and $\Ll$ constituents, which are $1/2^+$ and $1/2^-$ states, respectively.

Experience, from the deuteron onwards, indicates that molecular states are liable to form in systems where there is at least one S-wave channel, and where there is coupling to other channels in higher partial waves. As mentioned above, elastic one-pion exchange $\S\D^*-\S\D^*$ supports a bound state with $3/2^-$ quantum numbers, dominated by $\S\D^*$ in S-wave. The $\Ll\D$ pair also couples to $3/2^-$, but in P-wave. Including the \icc~ coupling can be expected to provide additional attraction in the $3/2^-$ state, which would then contain some admixture (however small) of $\Ll\D$ in P-wave.

But we may also expect an additional state in which the partial waves of $\SD$ and $\LD$ are interchanged, namely, with $\LD$ in S-wave and  $\SD$ in P-wave  \cite{Geng:2017hxc}. This state necessarily has $1/2^+$ quantum numbers,  which is a striking and unusual prediction of the model. States with positive parity seldom arise in molecular models, which typically involve S-wave constituents in a relative S-wave. In our model, the positive parity arises due to the P-wave nature of $\Ll$.

The prediction of $1/2^+$ and $3/2^-$ quantum numbers distinguishes our model from all other molecular scenarios, almost all of which predict $1/2^-$ and $3/2^-$. Likewise the hadro-charmonium model  predicts $1/2^-$ and $3/2^-$ quantum numbers for these states \cite{Eides:2015dtr,Eides:2017xnt,Eides:2019tgv}. Experimental measurement of the quantum numbers is therefore a key experimental test of our proposal.

With its unambiguous prediction of quantum numbers, our model is also quite different from most compact pentaquark scenarios, which typically allow for states with many possible quantum numbers, specifically  $1/2^-$, $3/2^-$ and $5/2^-$ for S-wave states, and $1/2^+$, $3/2^+$, $5/2^+$ and $7/2^+$ for P-waves~ \cite{Yuan:2012wz,Zhu:2015bba,Wang2015b,Maiani:2015vwa,Deng:2016rus,Chen:2016otp,Anisovich2015a,Ali:2016dkf,Wang:2019got,Weng:2019ynv,Zhu:2019iwm,Ali:2019npk,Holma:2019lxe,Giannuzzi:2019esi,Mutuk:2019snd,Cheng:2019obk}. (Although we note that some pentaquark scenarios predict a restricted spectrum, in particular models which incorporate the coupling to meson-baryon pairs \cite{Hiyama:2018ukv}, or which include additional dynamical assumptions \cite{Lebed:2015tna}.)

If it turns out that $P_c(4440)$ and $P_c(4457)$ have opposite parity, as predicted in our model, it would be very unnatural in compact pentaquark scenarios. Opposite parity implies a relative unit of orbital angular momentum, and the associated mass gap is expected to be much larger than the gap separating $P_c(4440)$ and $P_c(4457)$. A more general problem with the simplest compact multiquark scenarios is that they predict a vast number of states in apparent contradiction with data: this problem is particularly acute if the $P_c$ states have opposite parity, since the spectrum of the required P-wave multiplet is exceptionally rich.

\section{Binding}
\label{sec:binding}

In this section we give the pion exchange potentials relevant to the \icc~system, and solve the resulting Schr\"{o}dinger equation to obtain bound states consistent with $P_c(4440)$ and $P_c(4457)$.

\subsection{Potentials}

The one-pion exchange potential between heavy hadrons is derived from the momentum-space scattering amplitude, with vertices obtained from either the quark model or heavy-hadron chiral Lagrangians. In the static limit, both approaches give the same result, up to the overall normalisation and sign  \cite{bs}.  

If the constituent hadrons are considered as point particles, the potentials are singular at the origin. To account for the finite size of the hadrons -- and thus avoid the associated singular behaviour -- a phenomenological form factor is applied to each vertex, and this introduces some uncertainty into the calculation. We adopt a monopolar form factor at each vertex, using the same parametrisation and cut-off scale for all vertices, due to the similarity of the hadrons involved. The combination of two monopolar form factors results in a dipolar form factor
\begin{align}
F(\vq^2)&=\left(\frac{\Lambda^2-m^2}{\Lt^2+\vq^2}\right)^2,\\
\Lt^2&=\Lambda^2+\mu^2-m^2,
\label{eq:F}
\end{align}
where $\vq$ is the  pion three-momentum, and the ``recoil factor'' is \begin{equation}\mu^2=m^2-\omega^2,\label{eq:mu}\end{equation} where $m$  and $\omega$ are the pion mass and energy, respectively. Our results are not particularly sensitive to the parametrisation of the form factor, but are more sensitive to the chosen cut-off scale $\Lambda$, and we will comment on this below.

For the elastic \ecc~potential, we use the quark model,  in which all  vertices are obtained from the same basic quark-pion vertex. The resulting  momentum space potential,
\begin{align}
V(\vq) &=-\left(\frac{g_q}{f_\pi}\right)^2\frac{F(\vq^2)}{\mu^2+\q^2}\sum_{i}\vs_1^{(i)}\cdot\vq~\vs_2\cdot\vq~\vt_1^{(i)}\cdot\vt_2\label{eq:ope}
\end{align}
 is a generalisation of the familiar $NN$ potential, where here $\vs_{1}^{(1,2)}$ and $\vt_{1}^{(1,2)}$ are spin and isospin matrices acting on the two light quarks in  $\S$, and $\vs_2$ and $\vt_2$ act on the single light quark in $\D^*$.  We have used $g_q/f_\pi$  to parametrise the quark-pion coupling strength, following ref.~\cite{Thomas:2008ja}. In some other literature (such as \cite{Eides:2017xnt}) the quark axial coupling is used; it is related to $g_q$ by
 \begin{equation}g_q^A=-\sqrt 2g_q.\label{eq:axial}\end{equation}

The alternative approach, in which the vertices are derived from Lagrangians with heavy quark and chiral symmetries, is discussed in many papers. A key difference is that the resulting potentials are proportional to the product $gg_1$, where $g$ and $g_1$ are coupling constants associated with the meson and baryon vertices, respectively. Since the signs of these constants cannot be determined from experiment, the signs of the resulting potentials are ambiguous, which is problematic because these signs are critical in determining which channels support bound states. Some papers address this ambiguity by considering separately the possibilities that  $g$ and $g_1$ have the same or opposite signs. However, most authors ignore the ambiguity and assume (implicitly or explicitly) that the constants have the same sign.

This ambiguity is resolved in the quark model. Whilst the sign of $g_q$ cannot be obtained from experiment, it has no importance, since the potentials are proportional to $g_q^2$. For this reason, we prefer to work in the quark model approach.

For the inelastic \icc~potential, we use the heavy hadron chiral Lagrangian, as in ref.~\cite{Geng:2017hxc}. The momentum space potential is
\begin{equation}
V(\vq)=\frac{h_2g}{\sqrt 2 f_\pi^2}\frac{\omega F(\vq^2)}{\mu^2+\q^2}\bm{\epsilon}\cdot\vq~\mathbf{T}\cdot\vt\label{eq:ope2}
\end{equation}
where $\bm\epsilon$ is the polarisation vector of $\D^*$, $\mathbf T$ and $\vt$  are isospin matrices for the  $\S\Ll\pi$ and $\D^*\D\pi$ vertices, respectively, and $h_2$ and $g$ are the corresponding coupling constants. Although there is a sign ambiguity associated with the couplings $g$ and $h_2$, it has no impact on the results, since the associated potentials appear off the diagonal of the potential matrix.

The position space potentials are obtained by taking the Fourier transform of Eqs.~(\ref{eq:ope}) and~(\ref{eq:ope2}). Since molecular states are most likely for quantum numbers where there is at least one channel in S-wave, we concentrate on $J^P = 1/2^-$ and $3/2^-$ (where $\S\D^*$ is in S-wave), and $1/2^+$ (where $\LL\D$ is in S-wave). The one-pion exchange coupling of $\S\D^*$ and $\Ll\D$ to lower-lying decay channels will be considered perturbatively (Section~\ref{sec:opedecay}). 

\begin{table}
	\begin{tabular}{rccc}
		$\bm{J^P = 1/2^-}$	&	$\bigg|\begin{matrix}\S\D^*\\^2S_{1/2}\end{matrix}\bigg>$	&	$\bigg|\begin{matrix}\S\D^*\\^4D_{1/2}\end{matrix}\bigg>$&$\bigg|\begin{matrix}\LL\D\\^2P_{1/2}\end{matrix}\bigg>$ \\
		\noalign{\smallskip}
		$\bigg<\begin{matrix}\S\D^* \\^2S_{1/2}\end{matrix}\bigg|$	&	$4C$	&	$2\sqrt 2 T$	&	$W$	\\
		\noalign{\smallskip}
		$\bigg<\begin{matrix}\S\D^* \\^4D_{1/2}\end{matrix}\bigg|$	&	2$\sqrt 2 T$	&	$-2C+4T$	&	$-\sqrt 2 W$	\\
		\noalign{\smallskip}
		$\bigg<\begin{matrix}\LL\D\\^2P_{1/2}\end{matrix}\bigg|$	&	$W$	&	$-\sqrt 2 W$	&	0	\\
	\end{tabular}
	\\
	\vspace{0.5cm}
	\begin{tabular}{rcccc}
		$\bm{J^P = 3/2^-}$	&	$\bigg|\begin{matrix}\S\D^*\\^4S_{3/2}\end{matrix}\bigg>$	&	$\bigg|\begin{matrix}\S\D^*\\^2D_{3/2}\end{matrix}\bigg>$	&	$\bigg|\begin{matrix}\S\D^*\\^4D_{3/2}\end{matrix}\bigg>$	&	$\bigg|\begin{matrix}\LL\D\\^2P_{3/2}\end{matrix}\bigg>$	\\
		$\bigg<\begin{matrix}\S\D^*\\^4S_{3/2}\end{matrix}\bigg|$	&	$-2C$	&	$-2T$	&	$-4T$	&	$W$	\\
		\noalign{\smallskip}
		$\bigg<\begin{matrix}\S\D^*\\^2D_{3/2}\end{matrix}\bigg|$	&	$-2T$	&	$4C$	&	$2T$	&	$W$	\\
		\noalign{\smallskip}
		$\bigg<\begin{matrix}\S\D^*\\^4D_{3/2}\end{matrix}\bigg|$	&	$-4T$	&	$2T$	&	$-2C$	&	$-W$	\\
		\noalign{\smallskip}
		$\bigg<\begin{matrix}\LL\D\\^2P_{3/2}\end{matrix}\bigg|$	&	$W$	&	$W$	&	$-W$	&	0	\\
	\end{tabular}
	\\
	\vspace{0.5cm}\begin{tabular}{rccc}
		$\bm{J^P = 1/2^+}$	&	$\bigg|\begin{matrix}\S\D^*\\^2P_{1/2}\end{matrix}\bigg>$	&	$\bigg|\begin{matrix}\S\D^*\\^4P_{1/2}\end{matrix}\bigg>$	&	$\bigg|\begin{matrix}\LL\D\\^2S_{1/2}\end{matrix}\bigg>$	\\
		\noalign{\smallskip}
		$\bigg<\begin{matrix}\S\D^*\\^2P_{1/2}\end{matrix}\bigg|$	&	$4C$	&	$2\sqrt 2 T$	&	$W$	\\
		\noalign{\smallskip}
		$\bigg<\begin{matrix}\S\D^*\\^4P_{1/2}\end{matrix}\bigg|$	&	$2\sqrt 2 T$	&	$-2C+4T$	&	$-\sqrt 2 W$	\\
		\noalign{\smallskip}
		$\bigg<\begin{matrix}\LL\D\\^2S_{1/2}\end{matrix}\bigg|$	&	$W$	&	$-\sqrt 2 W$	&	0	\\
	\end{tabular}
	\caption{Pion exchange potentials for the \icc~system with $1/2^-$, $3/2^-$ and $1/2^+$ quantum numbers. The central ($C$), tensor ($T$), and vector ($W$) potentials are defined in the text.}
	\label{tab:potentials}
\end{table}

For total isospin equal to 1/2, reducing the spin-isospin-angular momentum degrees of freedom yields the potential matrices shown in Table~\ref{tab:potentials}, where $C(r)$, $T(r)$ and $W(r)$ are the central, tensor and vector potentials. We distinguish two possibilities for the central potential  ($C_0$ and $C_1$), as explained below. The potentials are given by:
\begin{align}
C_0(r)&=\frac{4g_q^2}{3}\frac{\mu^3}{12\pi f_\pi^2}\left[
Y(\mu r)-\frac{\Lt}{\mu}Y(\Lt r)-\frac{\Lambda^2-m^2}{2\mu\Lt}e^{-\Lt r}\right]\\
C_1(r)&=\frac{4g_q^2}{3}\frac{\mu^3}{12\pi f_\pi^2}\left[
Y(\mu r)-\frac{\Lt}{\mu} Y(\Lt r)-\frac{\Lt(\Lambda^2-m^2)}{2\mu^3}e^{-\Lt r}\right]
\label{eq:Cs}\\
T(r)&=\frac{4g_q^2}{3}\frac{\mu^3}{12\pi f_\pi^2}\left[
H(\mu r) -\frac{\Lt^3}{\mu^3} H(\Lt r) \right.\nonumber\\&\qquad\qquad\qquad\left.-\frac{\Lt(\Lambda^2-m^2)}{2\mu^3}(1+\Lt r)Y(\Lt r)\right]
\label{eq:T}\\
W(r)&={g h_2}\frac{\mu^2\omega}{2^{5/2}\pi f_\pi^2}\left[G(\mu r)-\frac{\tilde\Lambda^2}{\mu^2}G(\tilde\Lambda r)-\frac{\Lambda^2-m^2}{2\mu^2}e^{-\tilde\Lambda r}\right]
\end{align}
where
\begin{align}
Y(x)&=\frac{e^{-x}}{x},\\
G(x)&=\left(1+\frac{1}{x}\right)Y(x),\\
H(x)&=\left(1+\frac{3}{x}+\frac{3}{x^2}\right)Y(x).
\end{align} 

The central potentials are distinguished by the inclusion or exclusion of a term whose origin, in the unregulated potential, is a delta function centred at the origin: $C_1$ includes this term, $C_0$ does not. (The momentum-space potential (\ref{eq:ope}) corresponds to $C_1$ in position space; the momentum-space potential corresponding to $C_0$ can be found in ref.~\cite{Thomas:2008ja}, for example.)

Many authors do not include the delta function term, on the basis that the physical picture of pion-exchange between hadrons is well-motivated only at long distance; at short distance, where the wavefunctions of the different hadrons overlap, interactions among their quark constituents cannot justifiably be ignored. 
\begin{figure}
	\includegraphics[width=0.49\textwidth]{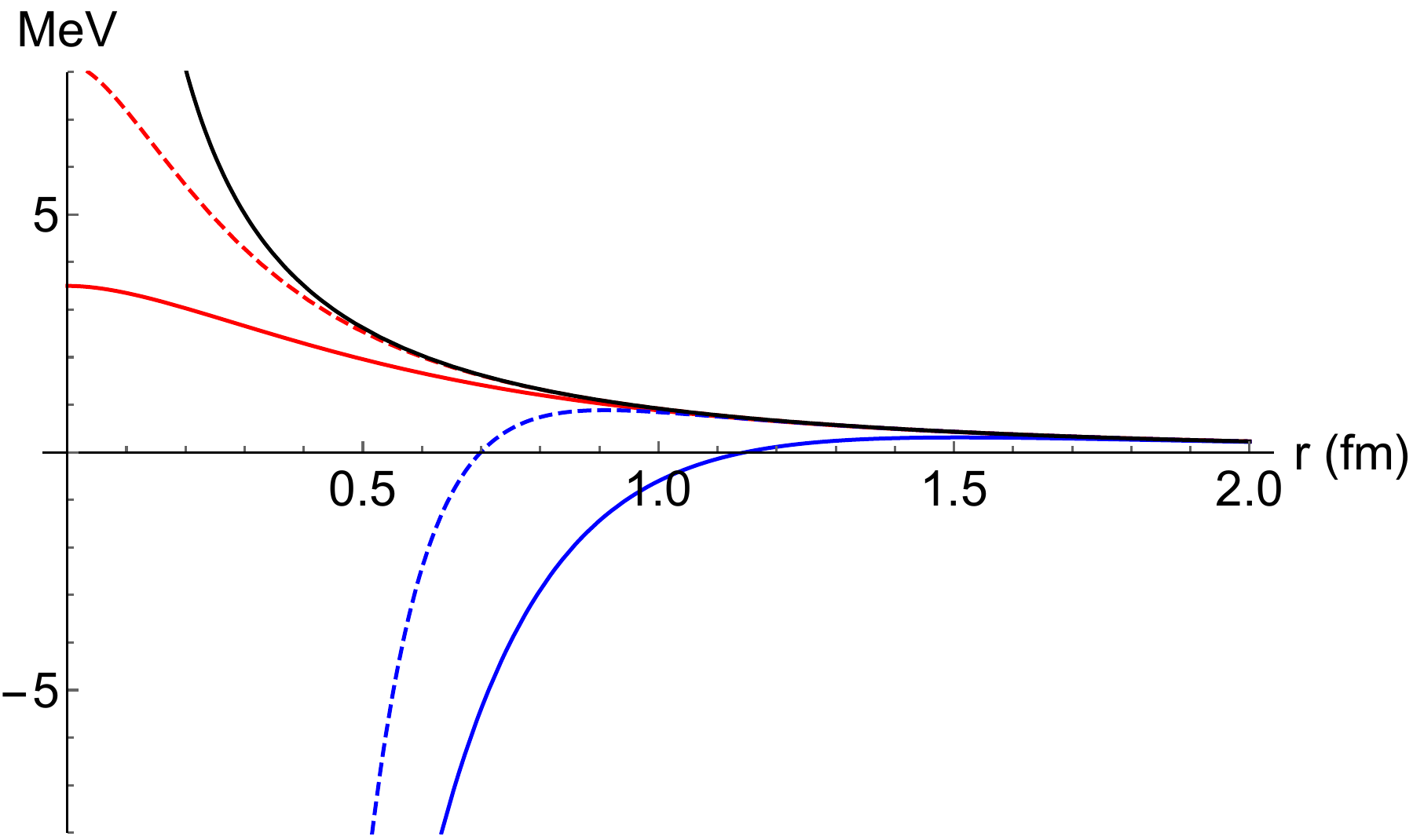}
	\caption{The central potentials $C_0(r)$ (red) and $C_1(r)$ (blue), with $\Lambda=1.0$~GeV (solid) and  $\Lambda=2.0$~GeV (dashed), and the un-regulated potential  (black).}
	\label{fig:pot}
\end{figure}

We plot the two central potentials in Figure~\ref{fig:pot}.
Because the delta function term has opposite sign to the long-distance Yukawa term, it has a drastic effect on the potential: notice that $C_0(r)$ (shown in red) has the same sign for all $r$, whereas $C_1(r)$ (blue) changes sign and becomes attractive at short-distance \cite{Thomas:2008ja,Liu:2019zvb}. Obviously, the predictions of the model will vary significantly depending on which of $C_0(r)$ or $C_1(r)$ is used, but it is surprisingly common in the literature to choose one or the other, without comment. Below we argue in favour of $C_0$.

All potentials (central, tensor and vector) become stronger as $\Lambda$ increases (see Fig.~\ref{fig:pot} for the central case). Consequently, in a typical calculation with S- and D-wave channels coupled by central and tensor interactions, binding is possible in most systems, provided $\Lambda$ is made large enough. The required value of $\Lambda$ varies significantly for different systems, and we identify the systems which bind with smaller values of $\Lambda$ as being most likely to support molecular states. The pattern of which states bind most easily is strongly influenced by the sign and magnitude of the central potential in the S-wave channel within each matrix \cite{Tornqvist:1991ks,Tornqvist:1993ng}, and by the chosen form ($C_0$ or $C_1$) of the central potential. 

The potential without the delta function term ($C_0$, shown in red in Fig.~\ref{fig:pot}) has the same sign everywhere, so the systems which bind most easily are those (such as our $3/2^-$ system) in which the S-wave central potential comes with negative sign. On the other hand, the potential which includes the delta function term ($C_1$, blue) has regions of both attraction and repulsion, but the short-distance behaviour clearly dominates. So the systems which bind most easily are those with an  attractive core to the potential -- namely, those in which the S-wave central potential which come with positive sign (such as our $1/2^-$ system). But this is conceptually problematic, because at long distance, where the potential is well-motivated theoretically, the corresponding potentials are repulsive. In this sense the  $C_1$ potential suffers from a problem of self-consistency: its predictions, which are driven by ambiguous short-distance behaviour, are in direct contradiction with the expectations derived from the more reliable long-distance behaviour. 

For both forms of the central potential, the binding energy increases monotonically with $\Lambda$, but the rate of increase is particularly rapid for the $C_1$ potential, which leads to a problem of fine tuning. A common motivation for molecular models is the existence of states with masses closely aligned to thresholds. But with the $C_1$ potential, due to the rapid increase in binding energy with $\Lambda$, shallow bound states only exist within a small parameter range. By comparison, with the $C_0$ potential the sensitivity to $\Lambda$ is weaker, and shallow bound states exist within a larger parameter range.

 In the absence of fine tuning, the $C_1$ potential typically predicts the existence of deeply bound states, in apparent contradiction with experimental data. We find that this is quite general, and suggest that this model by abandoned \cite{bs}. Related arguments against the $C_1$ potential can be found in refs.~\cite{Liu:2019zvb,Eides:2017xnt}. For most of our results we will use the $C_0$ potential, which we regard as preferable to the $C_1$ potential.
 
The modern effective field theory approach to this issue is to recognise that unknown short range dynamics exists, cut off the pion-exchange potential, and add cut-off-dependent short range interactions with strengths that are fit to data. When applied to models of the $P_c$ states based on $\S\*\D\*$ degrees of freedom \cite{Liu:2018zzu,Liu:2019tjn,Valderrama:2019chc,Meng:2019ilv,Sakai:2019qph}, a simultaneous fit to both $P_c(4440)$ and $P_c(4457)$ requires strong short-range interactions. In contrast, in our model we find that by including the the $\Ll\D$ channel, it is possible to fit both states without the need for additional short-range interactions.


For the elastic \ecc~couplings, there is considerable disagreement in the literature concerning the signs and magnitudes of the one-pion exchange potentials. Our results are identical to those of ref.~\cite{Eides:2017xnt} which, as far as we are aware, is the only other paper which obtains these potentials in the quark model (and which is thus free from sign ambiguities). We have also computed the corresponding potentials using heavy hadron chiral Lagrangians, and our results agree in magnitude with the those of refs~\cite{Yang:2011wz,Chen:2015loa,Chen:2016heh,Shimizu:2016rrd,Valderrama:2019chc}; those of refs~\cite{Shimizu:2017xrg,Shimizu:2018ran} are larger by a factor of 2. In terms of the signs,  our potentials agree with those of refs~\cite{Yang:2011wz,Shimizu:2016rrd,Shimizu:2017xrg,Shimizu:2018ran,Valderrama:2019chc}, provided $g$ and $g_1$ are chosen to have the same sign, as is done in most of those papers. By comparison, the potentials in refs \cite{Chen:2015loa,Chen:2016heh} have opposite sign to all the others -- again, assuming their chosen signs for $g$ and $g_1$.

For the inelastic \icc~potentials, our expressions (with the dipole form factor) agree with Geng~\etal\cite{Geng:2017hxc} once the pointlike constituent limit has been taken.

\subsection{Elastic channels}
\label{sec:elastic}

We begin by showing that if we switch off the $\LL\D$ channel, and thus have only the elastic \ecc~couplings, the model cannot simultaneously explain both $P_c(4440)$ and $P_c(4457)$. In particular, the common assignment of  $1/2^-$ and $3/2^-$ (in either order) does not work.

We solve the Schr\"{o}dinger equation, taking  $\mu=m$ for all potentials, and fixing the strength of the potentials by comparison to $NN$ scattering parameters \cite{Tornqvist:1993ng,Thomas:2008ja}, with 
\begin{align}
\frac{m^3g_q^2}{12\pi f_\pi^2}=1.3~\textrm{MeV},
\end{align}
where the pion decay constant is $f_\pi=132$~MeV.
The resulting value of $g_q\approx 0.6$ is equivalent, using (\ref{eq:axial}), to a value for the quark axial coupling constant $|g_q^A|$ which lies between the values $g_q^A=0.75$ and $g_q^A=1.0$ which are typically adopted in the literature~\cite{Yan:1992gz,Liu:2011xc,Eides:2017xnt}.

We find, with either the $C_0$ or $C_1$ potential, that bound states are possible for both $1/2^-$ and $3/2^-$ provided $\Lambda$ is made sufficiently large. By varying $\Lambda$ we may tune the masses of either state to match either of  $P_c(4440)$ or $P_c(4457)$. The problem is that the required values of $\Lambda$ differ considerably for the two quantum number channels, and this cannot be justified: the vertices in both channels are identical, and thus the same cut-off must be used.

The reason for this tension is that, as mentioned previously, the pattern of binding is intrinsically linked to the sign of the central potential in the S-wave channel within each matrix, and as shown in Table~\ref{tab:potentials}, the relevant matrix elements in the $1/2^-$ and $3/2^-$ systems have opposite sign.  Although our potentials are derived with the specific vertex model shown in equation~(\ref{eq:ope}), the relative signs and magnitudes of the central potentials are the same with other models for the vertex, such as the $^3P_0$ model or flux tube model~\cite{Burns:2014zfa}. Hence the problem we identify is rather general in nature.

The pattern of binding is correlated with the choice of potential in the manner described previously: with  the $C_0$ ($C_1$) potential, the systems which bind most easily are those in which the S-wave central potential comes with negative (positive) sign. Hence we find, for the $C_0$ potential, that $3/2^-$ binds most easily (with a smaller $\Lambda$), and in order to obtain the $1/2^-$ state, the cut-off must be pushed to a much larger value, which in turn renders the $3/2^-$ too deeply bound to fit with experimental data. (Recall that the same $\Lambda$ should be used for both channels, and that binding energies increase with~$\Lambda$.) The pattern reverses with the $C_1$ potential, but the essential  problem is the same: $1/2^-$ binds with a lower cut-off, and in order to obtain the $3/2^-$ state, the cut-off must be increased to such a value that the $1/2^-$ state becomes too deeply bound. 

An additional problem with both potentials is that the higher cut-off required to produce two $\S\D^*$ bound states ($1/2^-$, $3/2^-$) also implies that a further three $\S^*\D^*$ states ($1/2^-$, $3/2^-$, $5/2^-$) must also bind, and currently there is no experimental evidence for such states.

We thus conclude that the assignment of $P_c(4440)$ and $P_c(4457)$ as $1/2^-$ and $3/2^-$ states (or vice versa) is \textit{not} tenable in the simplest molecular scenario, where binding is due to the elastic \ecc~one-pion exchange coupling. The assignment can be made to work if additional, short-range interactions are included in the potential through contact terms fit to data \cite{Liu:2018zzu,Liu:2019tjn,Valderrama:2019chc,Peng:2019wys,Meng:2019ilv,Pan:2019skd,Sakai:2019qph},  exchange of mesons other than pions \cite{Xiao:2013yca,Xiao:2019aya,Xiao:2019gjd,Chen:2019asm,He:2015cea,He:2019ify,Liu:2019zvb}, or quark-level interactions \cite{Yamaguchi:2017zmn,Yamaguchi:2019seo}. These approaches typically predict the existence of a number of other states which are apparently not observed in experiment, in particular $\S^*\D^*$ states with $1/2^-$, $3/2^-$ and $5/2^-$ quantum numbers. The absence of the $5/2^-$ state in $J/\psi\, p$ may be understood due to its D-wave decay \cite{Shimizu:2019ptd,Takeuchi:2016ejt,Xiao:2013yca,Xiao:2019aya}, but there is no analogous argument for $1/2^-$ or $3/2^-$ states.

\subsection{Coupled-channels}

We now consider the full problem, including the $\Ll\D$ channel. 

There is some ambiguity in the inelastic \icc~potential, which is a consequence of working, as is customary, in the static limit. For inelastic transitions, the static limit is not compatible with energy conservation through the scattering diagram, hence the vertices $\Ll\S\pi$ and $\D\D^*\pi$ imply different values for the pion energy $\omega$, which appears in the vector potential $W$, and which also defines the effective mass $\mu$, through equation~(\ref{eq:mu}). The same ambiguity applies to  any calculation with static potentials for inelastic transitions, although it is almost never discussed in the literature. Typically the prescription $\mu=m$ is used for all potentials. 

These issues are discussed at length in Ref.~\cite{bs}; in this paper, we adopt a pragmatic approach of experimenting with different possibilities. The value of $\omega$ appearing in $W(r)$ is not so important, since we will anyway explore some variation in the overall strength of this potential. We define the combined coupling $\hat g$, where
\begin{equation}
\hat g \equiv \frac{g h_2 \omega}{2^{5/2} \pi f_\pi^2}.
\end{equation}
If we take
\begin{equation}
\omega = M(\LL) - M(\Sigma_c) = 139.4(5) \textrm{MeV}
\end{equation}
along with the couplings
\begin{equation}
g = 0.59(1)(7)
\end{equation}
obtained from $D^* \to D\pi$ and $D^* \to D \gamma$, and
\begin{equation}
h_2 = 0.62(8)
\end{equation}
obtained from $\LL \to \Sigma_c \pi$~\cite{Geng:2017hxc}, the preferred value of the combined coupling is
\begin{equation}
\hat g = 0.17 (4)\  \textrm{1/GeV}.
\end{equation}

For $\mu$ we should, strictly speaking, adopt an imaginary value, as determined by equation~(\ref{eq:mu}), and thus take the real part of the resulting complex potential $W(r)$. In practice, because the magnitude of the imaginary $\mu$ is tiny, the resulting potential is almost identical to that with a small, real $\mu$. (Oscillatory behaviour only appears at long distance, and is totally negligible.) Moreover, the potential hardly varies for real values of $\mu$ ranging from 0.001~GeV to the pion mass $m$, and we test both extremes below.

\begin{figure*}[ht]
\includegraphics[width=0.6\textwidth]{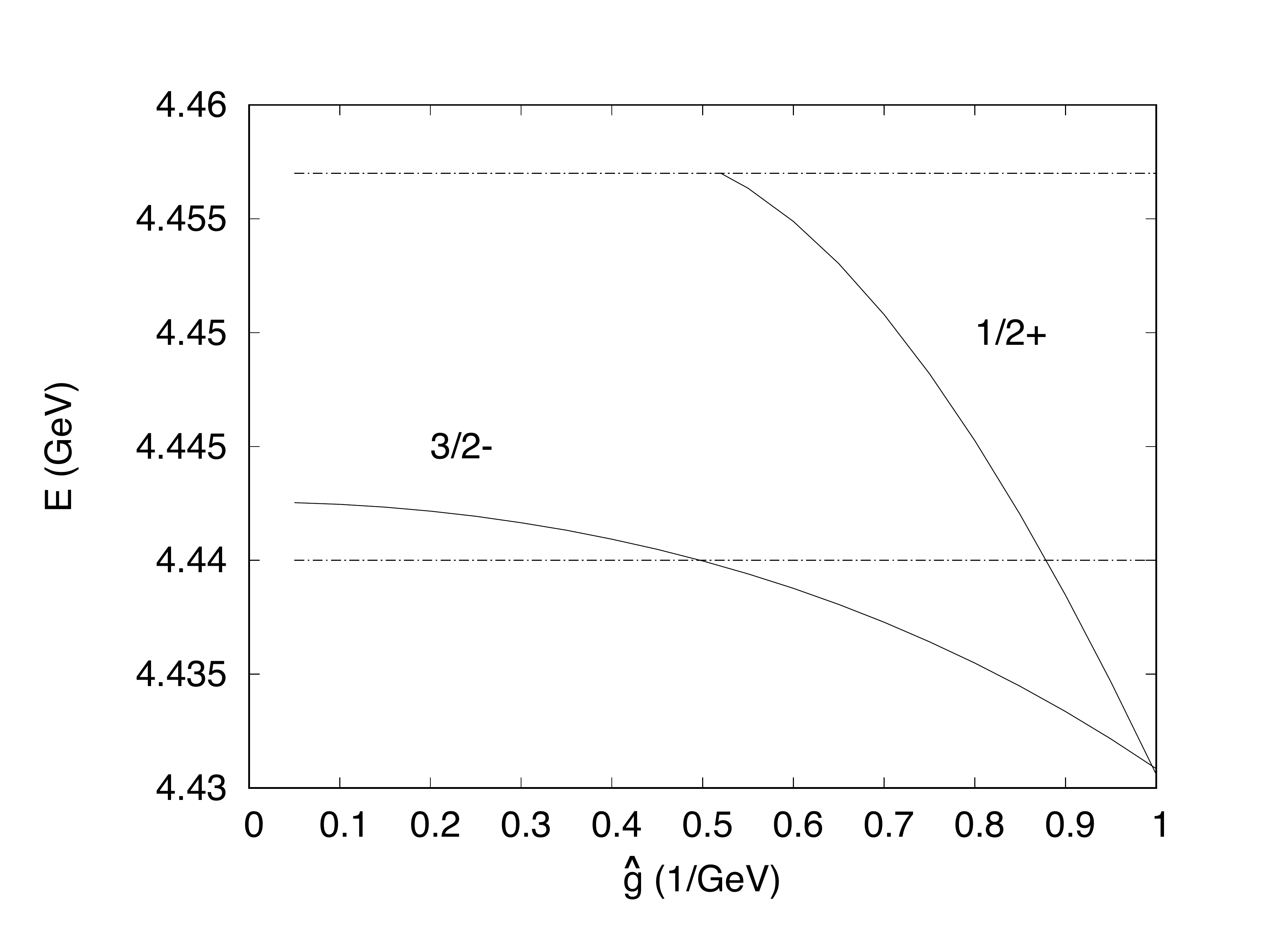}
\caption{Bound State Energy vs $\hat g$ for $\Lambda = 1.42$ GeV and $\mu = m$.}
\label{fig:g}
\end{figure*}

We solve the Schr\"{o}dinger equation with the potential matrices in Table \ref{tab:potentials}, 
and search for bound states while varying the cut-off $\Lambda$ and the coupling $\hat g$. As mentioned, we have found that using the central potential with regulated delta function leads to unreasonable phenomenology (and dubious self-consistency), we therefore focus on the option with no delta function ($C_0$ in Eq. \ref{eq:Cs}). In this case we find that the $J^P=1/2^+$ state is consistently higher in mass than the $3/2^-$. Since the higher mass $P_c$ state coincides with the $\Sigma_c\bar D^*/\Lambda_c(2595)\bar D$ threshold we adjust the cut-off until this state becomes virtual. We then adjust the combined coupling $\hat g$ to set the lower $3/2^-$ state at 4.440 GeV; see Fig. \ref{fig:g}. The parameters obtained in this way are $\Lambda = 1.42$ GeV and $\hat g =0.52$ GeV$^{-1}$.  As expected, the cut-off is somewhat larger than that typically used in nuclear physics (0.8 - 1.4 GeV), and is comparable to that required to create a weakly bound $X(3872)$ in a similar pion-exchange model~\cite{Swanson:2003tb}. 


Since the $1/2^+$ state is marginally bound its RMS radius is very large and it is dominated by the $\Lambda_c(2595)\bar D$ channel. Alternatively, the $3/2^-$ state has a relatively small RMS radius of 1.1 fm and channel components of  $\Sigma_c \bar D^*({}^2D_{3/2}) = 1.6\%$, $\Sigma_c \bar D^*({}^4S_{3/2}) = 85.7\%$, $\Sigma_c \bar D^*({}^4D_{3/2}) = 10.3\%$, and $\Lambda_c(2595) \bar D({}^2P_{3/2}) = 2.4\%$. No other bound states, including for isospin 3/2 or for $J^P = 1/2^-$, are found. Finally, the quoted results are obtained using $\mu=0.135$ GeV. We have found that these are insensitive to the value of $\mu$; for example using $\mu = 0.001$~GeV changes results at the level of a percent. 

The value of $\hat g$ required to fit the data is somewhat larger than the value preferred by $\Ll$ decays, but we do not consider this a problem. We note, for example, that including couplings to additional channels will change the preferred value of $\Lambda$, and in turn, $\hat g$. More importantly, we emphasise that, unlike other models, our fit has not required the introduction of additional \textit{ad hoc} short-range interactions which are fit to data.  We could of course add such terms to our potential, but at the cost of introducing additional free parameters. In this respect our model  is quite different to others in the literature, in which a simultaneous fit to both $P_c(4440)$ and $P_c(4457)$ relies crucially on the introduction of  strong short-range interactions which are fit to data~ \cite{Liu:2018zzu,Liu:2019tjn,Valderrama:2019chc,Meng:2019ilv,Sakai:2019qph}.

%
%
%
%
%
%

\section{Decays}
\label{sec:decays}

The LHCb collaboration has measured the widths of the higher mass $P_c$ signals as~\cite{Aaij:2019vzc}
\begin{align}
\Gamma[P_c(4457)] &= 6.4 \pm 2.0 {}^{+ 5.7}_{-1.9} \textrm{ MeV}, \\
\Gamma[P_c(4440)] &= 20.6 \pm 4.9 {}^{+8.7}_{-10.1} \textrm{ MeV}.
\end{align}

We seek to compute the widths of these states assuming the validity of the  molecular interpretation given above.  The relative complexity of molecular states permits a variety of possible decay mechanisms. Here we consider dissociative decays in which one of the bound hadrons undergoes a strong decay, short range interactions with a quark exchange component that can lead to hidden charm decay modes (which we call rearrangement decays), and long range pion-exchange-mediated decays.

\subsection{Dissociation Decays}


When sufficient phase space is available molecular states can decay by the strong decay of their constituents~\cite{Burns:2015dwa,Shen:2016tzq,Voloshin:2019aut}. 
An electroweak version of this mechanism would occur via neutron decay if the deuteron binding energy were less than the neutron-proton mass difference.
In the weak binding limit the width is simply given by the free space width weighted by the component fraction of the decaying hadron. In the case of the $1/2^+$ and $3/2^-$ resonances all the constituent hadrons are narrow: the $D^*$ has a width of 83 keV, the $\LL$ has a width of 2.6 MeV, and the $\Sigma_c$ width is approximately 1.8~MeV. 

Our model gives a sharp prediction for dissociation decays, which can be tested in experiment: $P_c(4457)$ decays to $\S\D^0\pi$, whereas $P_c(4440)$ decays to $\Lc\D^*\pi$. This arises from a combination of kinematics and isospin symmetry. The $\Ll\D^0$ combination, which dominates the wavefunction of $P_c(4457)$, couples only to $\S\D^0\pi$. The $\S\D^*$ combination couples to both $\S\D\pi$ and $\Lc\D^*\pi$, but for  $P_c(4440)$ in which this component dominates, only the latter mode is kinematically accessible. 

Using the above component fractions and constituent widths we estimate\begin{eqnarray}
\Gamma_{dis}[P_c(4457)\to\S\D^0\pi] &\approx&  2.6 \ \textrm{MeV} \nonumber \\
\Gamma_{dis}[P_c(4440)\to\Lc\D^*\pi] &\approx&  1.8 \ \textrm{MeV} 
\end{eqnarray}

\subsection{Pion-exchange Mediated Decays}
\label{sec:opedecay}

We now consider decays where pion exchange couples a constituent channel to a lighter hadronic final state. This mechanism has been discussed in refs~\cite{Eides:2018lqg,Lin:2017mtz}.

In our model $P_c(4457)$ is narrow because its dominant $\LL \bar D$ component does not directly couple to any lighter hadronic channel via perturbative one pion exchange. On the other hand, $P_c(4440)$ is broad because its dominant $\S\D^*$ component couples to $\Lambda_c\bar D$, $\Lambda_c \bar D^*$, and $\Sigma_c\bar D$. One might expect the S-wave $\Sigma_c\bar D^*$ component to dominate these processes; however, the strong tensor interaction present in the system implies that D-waves must also be considered. 

If one assumes point-like hadrons then the decay amplitude is the overlap of a molecular wavefunction component with the central or tensor potential projected onto a given wave. Specifically,
the partial width to a given final state with angular momentum $L$ is given by
\begin{equation}
\Gamma_L(k) =  \frac{k}{4\pi^2} \frac{E_B E_C}{m_{P_c}} |\mathcal{A}_L(k)|^2
\end{equation}
where

\begin{equation}
\mathcal{A}_0(k) = 4\pi \sum_\alpha \int rdr \,[u_{\alpha|S} V_C(r) + u_{\alpha|D}V_T(r)] j_0(kr)
\end{equation}
and
\begin{equation}
\mathcal{A}_2(k) = 4\pi \sum_\alpha \int rdr \,[u_{\alpha|S} V_T(r) + u_{\alpha|D}V_C(r)] j_2(kr).
\end{equation}
The index $\alpha$ refers to a molecular wavefunction component, while $\alpha|\ell$ is a component with angular momentum $\ell$. The interactions $V_C$ and $V_T$ consist of a channel coupling coefficient times the central or tensor interactions shown in Eqs. (\ref{eq:Cs}) and (\ref{eq:T}). These coefficients are shown in Table~\ref{tab:opedecays}.

\begin{table}
	\begin{tabular}{rccc}
&	$\left|\begin{matrix}\S\D^*\\^4S_{3/2}\end{matrix}\right>$	&	$\left|\begin{matrix}\S\D^*\\^2D_{3/2}\end{matrix}\right>$	&	$\left|\begin{matrix}\S\D^*\\^4D_{3/2}\end{matrix}\right>$	\\	
\noalign{\smallskip}
$\left<\begin{matrix}\Lc\D\\^2D_{3/2}\end{matrix}\right|$&$3\sqrt 3T/2$&$3\sqrt 3C/2$&$-3\sqrt 3T/2$\\
\noalign{\smallskip}
$\left<\begin{matrix}\S\D\\^2D_{3/2}\end{matrix}\right|$&$-2\sqrt 3 T$& $-2\sqrt 3 C$&$2\sqrt 3 T$\\
\noalign{\smallskip}
$\left<\begin{matrix}\Lc\D^*\\^4S_{3/2}\end{matrix}\right|$&$3C/2$&$3T/2$&$3T$\\
\noalign{\smallskip}
$\left<\begin{matrix}\Lc\D^*\\^2D_{3/2}\end{matrix}\right|$&$3T/2$&$-3C$&$-3T/2$\\
\noalign{\smallskip}
$\left<\begin{matrix}\Lc\D^*\\^4D_{3/2}\end{matrix}\right|$&$3T$&$-3T/2$&$3C/2	$\\
\noalign{\smallskip}
$\left<\begin{matrix}\S^*\D\\^4S_{3/2}\end{matrix}\right|$&$-\sqrt 3C$&$-\sqrt 3 T$&$\sqrt 3	T$\\
\noalign{\smallskip}
$\left<\begin{matrix}\S^*\D\\^4D_{3/2}\end{matrix}\right|$&$\sqrt 3T$&$\sqrt 3 T$&$-\sqrt 3	C$
\end{tabular}

\caption{The channel coupling coefficients, as defined in the text, for decays mediated by one-pion exchange.}
\label{tab:opedecays}
\end{table}

Summing over the three $\Sigma_c$ components of the $3/2^-$ state $P_c(4440)$ gives
\begin{align}
\Gamma_{ope}[P_c(4440) \to \Lambda_c\bar D^*({}^4S)] &= 47\ \textrm{MeV}\\
\Gamma_{ope}[P_c(4440) \to \Lambda_c\bar D^*({}^2D)] &= 3\ \textrm{MeV}\\
\Gamma_{ope}[P_c(4440) \to \Lambda_c\bar D^*({}^4D)] &= 29\ \textrm{MeV}\\
\Gamma_{ope}[P_c(4440) \to \Lambda_c\bar D({}^2D)] &= 17\ \textrm{MeV}\\
\Gamma_{ope}[P_c(4440) \to \Sigma_c\bar D({}^2D)] &= 14\ \textrm{MeV}\\
\Gamma_{ope}[P_c(4440) \to \Sigma_c^*\bar D({}^4S)] &= 0.7\ \textrm{MeV}\\
\Gamma_{ope}[P_c(4440) \to \Sigma_c^*\bar D({}^4D)] &= 0.8\ \textrm{MeV.}
\end{align}
Hence the dominant mode is $\Lc\D^*$, consistent with results in other literature~\cite{Ortega:2016syt,Eides:2018lqg,Lin:2017mtz}.

\subsection{Rearrangement Decays}
\label{sec:rearr}

\begin{figure}
\includegraphics[width=0.4\textwidth]{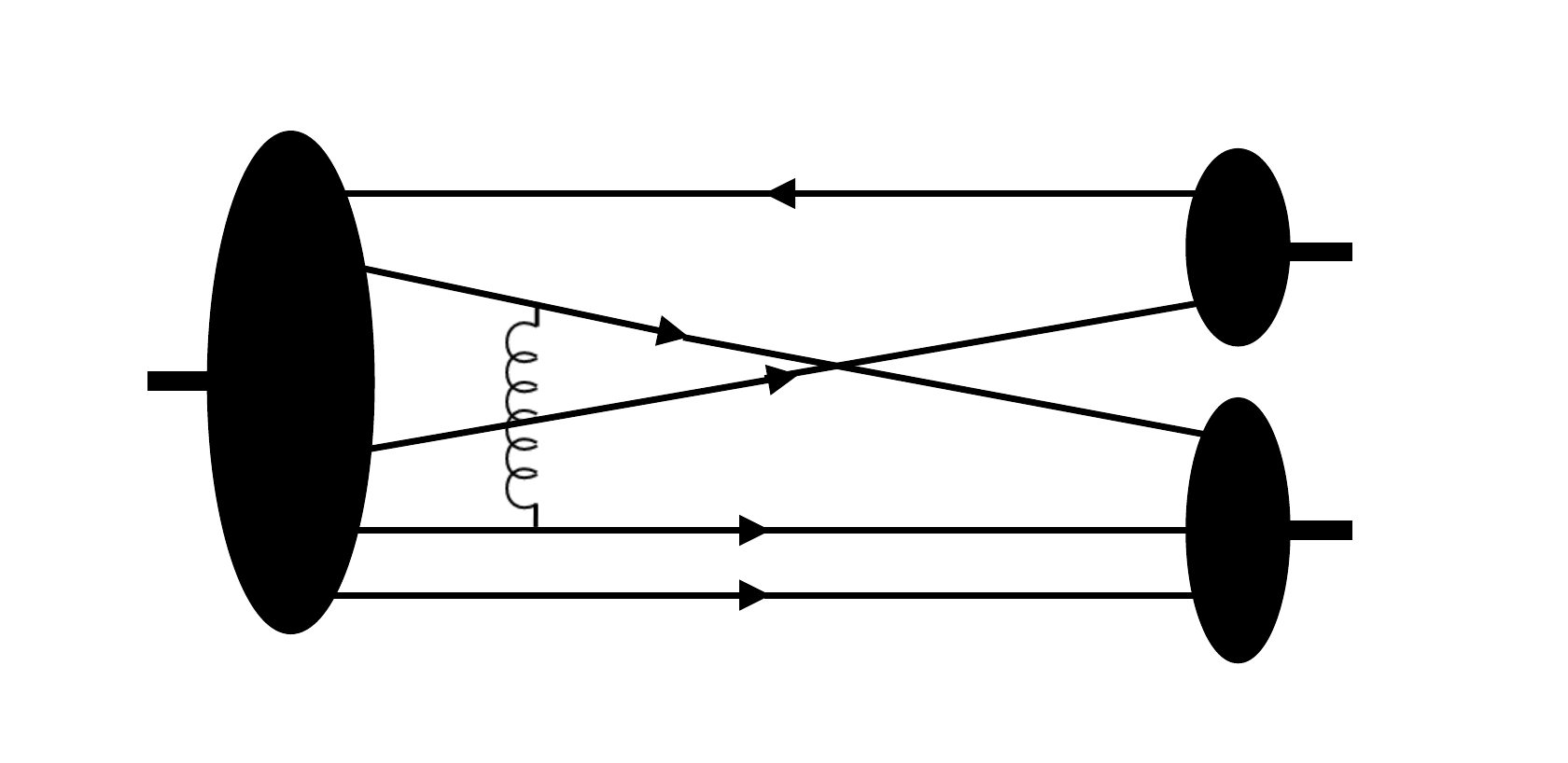}
\caption{A Diagram Contributing to the Rearrangement Decay Amplitude.}
\label{fig:rd}
\end{figure}

A novel decay mechanism available to hadronic bound states consists of the exchange of quarks and gluons between constituent hadrons which then yields a lower energy hadronic final state. For example, $\Sigma_c \bar D^*(3/2^-)$ can scatter into $J/\psi \, p$ in an S-wave while $\LL \bar D(1/2^+)$ goes to $J/\psi\, p$ in a P-wave. The expected partial wave suppression therefore suggests
\begin{equation}
\Gamma(P_c(4457)^+\to J/\psi\, p)<\Gamma(P_c(4440)^+\to J/\psi\, p).\label{eq:of}
\end{equation}

The formalism is very similar to that for pion exchange-mediated decays with the chief difference being the replacement of the OPE potential with the rearrangement scattering amplitude. In this case the decay amplitude can be written as 

\begin{equation}
{\mathcal A}_L(k) = \sum_{\alpha \ell} \int \frac{q dq}{(2\pi)^3}\,  T_{L,\ell}(k,q) u_{\alpha|\ell}(q)
\label{eq:re}
\end{equation}
where $u_{\alpha|\ell}$ is the radial hadronic bound state wavefunction as before and $T_{L,\ell}$ is the scattering amplitude from the bound state channel $\alpha$ with relative orbital momentum $\ell$ to the decay channel with angular momentum $L$. The T-matrix element can be estimated in the constituent quark model by computing the Born order scattering amplitude. This consists of one application of a quark model interaction followed by quark exchange, which is required to maintain colour neutrality. This general approach is reasonably reliable where it can be tested and has been applied to reactions of interest here in Ref. 
\cite{Hilbert:2007hc}. Typically inelastic cross sections range in strength from approximately 7 mb near threshold for $J/\psi\, p \to \bar D^0\Lambda_c$ to 0.1 mb for $J/\psi\, p \to \bar D^{0*} \Sigma_c^+$. Because the latter cross sections are quite small we expect small rearrangement decay widths. We therefore choose to implement a simple separable Ansatz for the S-wave T-matrix element, 

\begin{equation}
T(k,q) = a \exp(-k^2/b^2) \cdot a \exp(-q^2/b^2),
\end{equation}
and determine $a \approx 3$ GeV$^{-1}$ and $b \approx 0.8$ GeV in a fit to the computed expression for $J/\psi\, p \to \Sigma_c \bar D^*$ in $J=3/2$, $L=0$. As expected the resulting width obtained from Eq. (\ref{eq:re}) is small (approximately 1 keV), and therefore can be neglected. 

The GlueX experiment recently reported an upper limit on photo-production of the LHCb states off the proton, which implies a tight upper limit on their decay branching fraction to $J/\psi\, p$ \cite{Ali:2019lzf}. The same observation was made with reference to earlier photo-production data~\cite{Wang:2015jsa}. Our results are consistent with these upper limits.

With a model based on exchange of charmed hadrons, Eides~\etal~\cite{Eides:2018lqg} obtain a considerably larger $J/\psi\, p$ width of 30~keV, still totally insignificant on the scale of the measured total widths, and consistent with GlueX. Xiao~\etal~\cite{Xiao:2019mvs}, using  effective Lagrangians,  find $J/\psi\, p$ widths which   are orders of magnitude larger, on the scale of several MeV; these are not consistent with the GlueX limits~\cite{Sakai:2019qph}. Related earlier predictions gave even larger widths~\cite{Lu:2016nnt}.  QCD sum rule calculations for $P_c(4312)$~\cite{Wang:2019hyc,Xu:2019zme} also give large $J/\psi\,p$ widths, inconsistent with GlueX.

\subsection{Remarks}

Given the GlueX upper limits on the observed $J/\psi\, p$ mode, the $P_c$ states must decay dominantly into modes which have not yet been seen in experiment. For $P_c(4440)$ we predict that the dominant decay modes are to $\Lambda_c \bar D\*$ in various waves, and to $\Sigma_c\bar D$. Furthermore, D-waves are not suppressed with respect to S-waves because of the strong tensor forces present. On the other hand, for $P_c(4457)$ we predict the dominance of the three-body mode $\S\D^0\pi$ due to the dissociation decay of the $\Ll$ constituent.

Our final estimates for the widths are therefore 
\begin{eqnarray}
\Gamma_{tot}[P_c(4457)] &\approx&  3  \ \textrm{MeV} \nonumber \\
\Gamma_{tot}[P_c(4440)] &\approx&  111 \ \textrm{MeV} .
\end{eqnarray}
Although these results correctly reproduce $\Gamma(4457) < \Gamma(4440)$ and obtain a small width for the $P_c(4457)$, the predicted total width for the $P_c(4440)$ is too large. We comment on this in the Conclusions.

\section{Isospin}
\label{sec:isospin}

After the initial discovery of $P_c(4380)$ and $P_c(4450)$, one of us made the observation that, in the molecular scenario based on $\S^*\D$ and $\S\D^*$ degrees of freedom, these states would be admixtures of $I=1/2$ and $I=3/2$~\cite{Burns:2015dwa}. The decays $J/\psi \Delta$ and $\eta_c \Delta$ were proposed as striking signatures of this isospin mixing, and it was argued that spin symmetry implies these modes could have significant branching fraction, even if the isospin mixing angles are small~\cite{Burns:2015dwa}.

Isospin mixing arises because the mass gap between the thresholds for the charge combinations $\S^{(*)+}\D^{(*)0}$ and $\S^{(*)++}D^{(*)-}$  is significant on the scale of the molecular binding energy. The mechanism is similar to the analogous mechanism for $X(3872)$ \cite{Close:2003sg,Swanson:2003tb}.

After the update from LHCb, $P_c(4457)$ is significantly closer to $\S^+\D^{0*}$ threshold than its predecessor $P_c(4450)$, which implies that the isospin mixing is likely to be even more prominent. This was recently quantified by Guo~\etal~\cite{Guo:2019fdo}, who find that if $P_c(4457)$ is a $\S\D^*$ molecule, isospin mixing is at the level of a few percent to around 30\%.

But as noted in ref.~\cite{Burns:2015dwa}, the isospin mixing is uniquely associated with $\S\*\D\*$ degrees of freedom, due to the presence of two charge combinations with different thresholds. For components with only one charge combination, such $\Ll\D^0$, there is no contribution to isospin mixing. For this reason, in our model we find that isospin mixing is very small: the $P_c(4457)$ is dominantly $\Ll\D$, and the only contribution to isospin mixing is from $\S\D^*$, which is a small component of the wavefunction. We thus predict, in contrast to other models, that the decays of $P_c(4457)$ to $J/\psi \Delta$ and $\eta_c \Delta$ will be negligible.

As for the $P_c(4440)$, we also find that isospin mixing is small. In this case, although there is  considerable $\S\D^*$ in the wavefunction, the binding energy is large compared to threshold mass gap, and isospin mixing is not significant.

Isospin was also considered by ref.~\cite{Guo:2019kdc} using the effective range expansion; they conclude that isospin mixing in all three of the new $P_c$ states is mild.

\section{Production}
\label{sec:production}

In this section, we make a general point which applies not only to $P_c(4440)$ and $P_c(4457)$, but also the lighter states $P_c(4312)$ and $P_c(4380)$. The vast majority of literature on these states describes them in terms of $\S\*\D\*$ degrees of freedom, but the question of how such configurations are produced in $\Lambda_b$ decays has not adequately been addressed. We will show that there is an outstanding issue here which needs to be resolved.

\begin{figure}
	\includegraphics[width=0.5\textwidth]{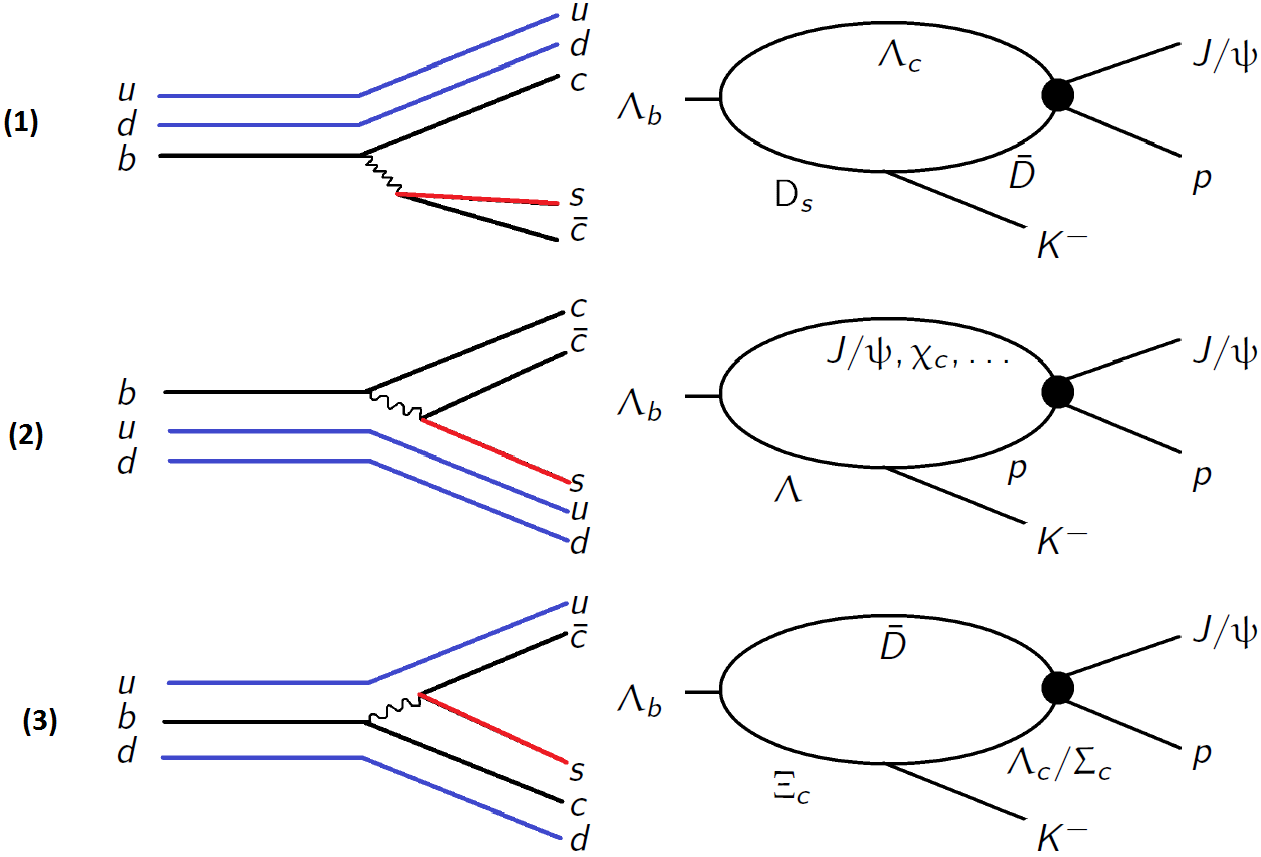}
	\caption{Diagrams showing the production of meson-baryon molecular states from $\Lambda_b$ decays. The left panel shows the three quark-level topologies for the weak decay of $\Lambda_b$: Topology~(1) is colour-favoured, while Topologies~(2) and (3) are colour-suppressed. The right panel shows the corresponding loop diagrams which produce the $J/\psi\, p K^-$ final state. (In these diagrams, the hadron labels refer to the flavour content, so ``$\Lc$'' indicates any of $\Lc$, $\Lc(2595)$ or $\Lc(2625)$, for example.)}
	\label{fig:prod}
\end{figure}

In Figure~\ref{fig:prod} we show the different production diagrams. The left panel shows the three possible topologies for the Cabibbo-favoured weak decay $(b\to cs\bar c)$, and the right panel shows the corresponding loops resulting in a $J/\psi\, p K^-$ final state. In each case, $K^-$ is produced from the loop, and the possible molecular constituents are those which feed the $J/\psi\, p$ final state (black circle). In these diagrams, the labels for the hadrons in the loop refer only to flavour, so that ``$\Lc$'' indicates any of $\Lc$, $\Lc(2595)$ or $\Lc(2625)$, for example, and ``$\D$'' can be $\D$ or $\D^*$.

As far as the weak vertex is concerned, Topology~(1) is dominant: its weak vertex  is colour-favoured, whereas the vertices in Topologies (2) and (3) are colour-suppressed. Since the $ud$ pair is a spectator in the weak transition, conservation of isospin and spin implies that this dominant diagram can only produce baryons with $\Lambda_c$ flavour, not $\S$ flavour. We therefore do not agree with the mechanisms discussed in refs~\cite{Mikhasenko2015,Wu:2019rog}, which feature $\S$ baryons produced in diagrams of this type. 

The $\S\*\D\*$ combinations favoured by most interpretations of the $P_c$ states can be produced via Topology~(3) of Figure~\ref{fig:prod}, and this mechanism involves a colour-suppressed weak decay. In this case the $u$ and $d$ quarks of $\Lambda_b$ end up in different hadrons, and the $\S\*$ baryon is produced through the strong decay of a baryon with $\Xi_c$ flavour in the loop. Note that, as well as producing $\S\*\D\*$ states, this topology also provides another mechanism for the production of $\Lc\D\*$ states. It is also the only phenomenologically relevant intrinsically non-factorisable weak decay of which we are aware.

We conclude that the production of the molecular components  $\S\*\D\*$ favoured by most models is suppressed in $\Lambda_b$ decays: they arise only due to isospin-breaking in the colour-favoured Topology (1), or due to the colour-suppressed Topology (3). 

This points to the importance of including coupled-channels such as $\S\*\D\*-\Lc\D\*$ and $\S\*\D\*-\Ll\D\*$ in any description of the $P_c$ states. In this picture, the states can be produced through their wavefunction components, such as $\Lc\D\*$ and $\Ll\D$, which are accessible via the colour-favoured mechanism. Even if the corresponding wavefunction components are comparatively small, the dominance of the colour-favoured mechanism (which we quantify below) implies these components are important in the production of $P_c$ states.

All of this argues in favour of the importance of one-pion exchange in the formation of the $P_c$ states. The $\S\*\D\*$ channels are undoubtedly important in the binding of these states, but only components with an isoscalar baryon, such as $\Lc\D\*$ and $\Ll\D$, are copiously produced in $\Lambda_b$ decays. Coupling between the channels can resolve this issue, and such coupling implies the exchange of pions, or more generally, isovector mesons. (We note in particular that the commonly used effective field theory approach, with only contact terms and $\S\*\D\*$ constituents, does not have the required channel coupling.)

Coupled channel effects are expected to be particularly important when the corresponding thresholds are close in mass, and it is interesting to note that the masses of all three of the new $P_c$ states are near to relevant threshold pairs. We have already argued for the relevance of  the \icc~coupling to the $P_c(4440)$ and $P_c(4457)$ states, but we also note that the other new state, $P_c(4312)$, is very close to a pair of nearby thresholds ($\Lc\D^*$ and $\S\D$) which couple via one-pion exchange, hence the mechanism we describe is also likely to be relevant there.

Quantifying the scale of the suppression of $\S\*\D\*$ configurations is very difficult, as it requires modelling both the electroweak vertex and the strong vertex in the loop, and summing over a large number of intermediate states where poorly constrained interference effects can be important. We may, however, estimate the scale of the suppression with reference to the electroweak part of the vertex. We assume that the isospin breaking mechanism is sub-dominant, so that $\S\*\D\*$ states are produced through Topology (3).

The colour-favoured diagram (1) is responsible for the largest observed \cite{Tanabashi:2018oca} two-body decay mode of $\Lambda_b$,
\begin{equation}
{\cal B}(\Lambda_b\to \Lc
 D_s^-)=(1.10\pm 0.1)\%,
\end{equation}
Such decays can be computed with reasonable accuracy by the heavy quark formalism. For example, for $\Lambda_b\to \Lc D_s^-$, the amplitude
\begin{equation}
i\mathcal{M} = \frac{G_F}{\sqrt{2}} V_{bc}V^*_{cs} i f_{D_s} p^\mu_{D_s} \xi(w) \bar u_c \gamma_\mu u_b
\end{equation}
(where $\xi$ is the Isgur-Wise function, $V_{bc}$ and $V_{cs}$ refer to CKM matrix elements, and the $D_s$ decay constant is approximately 260 MeV) 
gives
\begin{equation}
\Gamma(\Lambda_b \to \Lambda_cD_s^-) = 5.8\cdot 10^{-15} \ \textrm{GeV,}
\end{equation}
which is equivalent to
\begin{equation}
{\cal B}(\Lambda_b\to \Lc D_s^-)=1.3\label{eq:cf1}\% .
\end{equation}
A related prediction is \cite{Chua:2019yqh}:
\begin{align}
{\cal B}(\Lambda_b\to \Lc D_s^{*-})&=(1.830^{+0.849}_{-0.629})\label{eq:cf2}\%
\end{align}

As for the colour-suppressed $\Lambda_b\to \Xi_c^{(\prime,*)}\D\*$ transitions required to produce $\S\*\D\*$ molecular components, there is no experimental data, nor are there any theoretical predictions for the branching fractions, presumably because this topology is non-factorisable, and typically assumed to be small. (There are $SU(3)$ relations among the transitions~\cite{Arora:1992yq,Du:1994qt}, but we are not aware of any prediction for their magnitude.) The heavy quark formalism does not work in this case, as it is a non-factorisable decay. Hence we make a quark model computation, taking the $\Xi_c\D^*$ mode as an example. The amplitude is, suppressing spin indices:
\begin{eqnarray}
i\mathcal{M}(P) &=& \sqrt{2E_D}\sqrt{2 E_\Xi}\sqrt{2E_{\Lambda_b}} \frac{G_F}{\sqrt{2}} V_{bc}V^*_{cs}\frac{1}{N_c}  \nonumber \\
&& \times  \int \frac{d^3q}{(2\pi)^3} \, \frac{d^3k_2}{(2\pi)^3} \,\frac{d^3k_3}{(2\pi)^3} \nonumber \\
&& \times
\phi_{\Lambda_b}(-k_2-k_3,k_2,k_3)\cdot  \bar u_b(k_3) \Gamma^\mu u_c(k_3-q) \nonumber\\
&&\times\phi^*_{\Xi}(-k_2-k_3,q+k_2-P,k_3-q) \nonumber \\
&& \times \bar v_c(P-k_2) \Gamma_\mu u_s(q+k_2-P) \phi^*_D(P-k_2,k_2) \nonumber\\
\end{eqnarray}
where $\Gamma^\mu = \gamma^\mu - \gamma_5 \gamma^\mu$. The Fermi decay constant is denoted $G_F$ and is approximately $1.17\cdot 10^{-5}$ GeV$^{-2}$. This expression was evaluated in the leading nonrelativistic limit using simple Gaussian wavefunctions with a universal scale set to 
400 MeV. The wavefunction scale is typical for ground state hadrons involving light quarks~\cite{Barnes:1991em}.
The result is
\begin{equation}
\Gamma(\Lambda_b \to \Xi_c \bar D^*) \approx 1\cdot 10^{-16} \ \textrm{GeV},
\end{equation}
equivalent to
\begin{equation}
\mathcal B(\Lambda_b \to \Xi_c \bar D^*)=2.5\cdot 10^{-4}.
\label{eq:cs}
\end{equation}
We conclude, by comparison to equations~(\ref{eq:cf1}) and (\ref{eq:cf2}), that colour suppressed vertices are smaller by more than a factor of 50.

We now compare to experimental data and show that this level of suppression really is a problem for models based solely on $\S\*\D\*$ degrees of freedom.
The branching fractions $\mathcal B (\Lambda_b\to P_c K^-)$ are not known, although the LHCb collaboration has measured a ratio of branching fractions defined as 
\begin{equation}
{\mathcal R} = \frac{
{\mathcal B}(\Lambda_b^0 \to P_c^+ K^-) \, 
{\mathcal B}(P_c^+ \to J/\psi \,p) } {
{\mathcal B}(\Lambda_b^0 \to J/\psi \, p\, K^-) }.\label{eq:r}
\end{equation}
Values for this  ratio are reported as \cite{Aaij:2019vzc} 
\begin{equation}
\mathcal R=\left\{\begin{array}{ll}
(0.53 \pm 0.16 {}^{+0.15}_{-0.13})\%,&[P_c(4457)]\\
(1.11 \pm 0.33 {}^{+0.22}_{-0.10})\%,&[P_c(4440)]\\
(0.30 \pm 0.07 {}^{+0.34}_{-0.09})\%,&[P_c(4312)].
\end{array}\right.\label{eq:rats}
\end{equation}

\begin{figure}
	\includegraphics[width=0.5\textwidth]{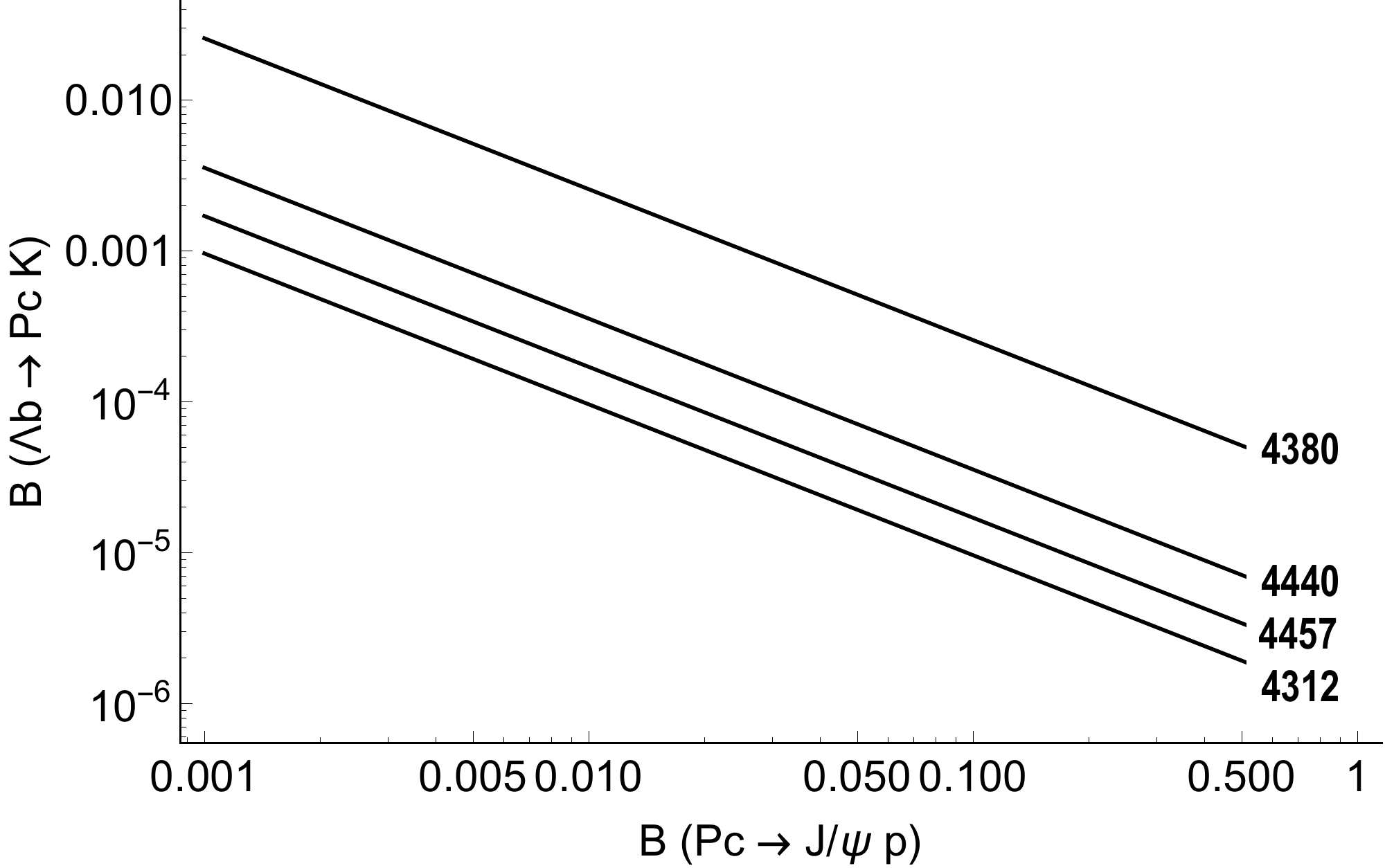}
	\caption{The branching fractions $\mathcal B (\Lambda_b^0\to P_c^+ K^-)$ as a function of $\mathcal B (P_c^+\to J/\psi\, p)$, obtained as described in the text.}
	\label{fig:bfs}
\end{figure}

Using the PDG value for the denominator, the product of branching fractions in the numerator can be computed for each of the three $P_c$ states~\cite{Cao:2019kst}. We show in Fig.~\ref{fig:bfs} the resulting relation between ${\mathcal B}(\Lambda_b^0 \to P_c^+\,K^-) $ and ${\mathcal B}(P_c^+ \to J/\psi \,p)$. We have also included $P_c(4380)$ in the plot, using the measured product of branching fractions from ref.~\cite{Aaij:2015tga}, although the existence of this state, and the product of  branching fractions, require confirmation following the latest LHCb results. The plot shows the central values only; we do not include errors, because the discussion below concerns the overall scale of the branching fractions, not their precise values.

Based on the absence of signals in their photo-production data, GlueX \cite{Ali:2019lzf} recently computed the following upper limits (at $90\%$ confidence level) on the $J/\psi  p$ branching fractions,
\begin{equation}
\mathcal B (P_c^+\to J/\psi\, p)<\left\{\begin{array}{ll}
3.8\%,&[P_c(4457)]\\
2.3\%,&[P_c(4440)]\\
4.6\%,&[P_c(4312)] 
\end{array}\right. .
\end{equation}
These limits are computed using a variant of the JPAC model~\cite{Blin:2016dlf}, and assume that the states have $3/2^-$ quantum numbers. (This assignment would be unusual for $P_c(4312)$, but as already discussed, is widely used for $P_c(4440/4457)$). We may expect some variation in these limits if other quantum numbers are assumed, but it is interesting to note that another computation, in an analysis of earlier photo-production data, arrived at similar limits for the precursor states $P_c(4450)$ and $P_c(4380)$ \cite{Wang:2015jsa}.

Referring to Fig.~\ref{fig:bfs}, the GlueX limits imply that lower limits on $\mathcal B (\Lambda_b^0\to P_c^+ K^-)$ are at the level of $10^{-4}$ for $P_c(4312/4440/4457)$, and $10^{-3}$ for $P_c(4380)$. These are implausibly large branching fractions if the molecular states are produced via colour-suppressed weak decay, given that the electroweak vertex is also at the level of $10^{-4}$: see eqn~(\ref{eq:cs}). 

By way of comparison, we note that the production of the $D^*\D$ molecular candidate $X(3872)$ in $B$ decays can proceed via a colour-favoured decay, in a diagram similar to Topology (1) in Fig.~\ref{fig:prod}, but with $\Lc D_s\*$ replaced by $D\* D_s\*$. In this case the production branching fraction
\begin{equation}
\mathcal B(B^+\to X(3872)K^+)<2.6\cdot 10^{-4},
\end{equation}
is at least a factor of 50 smaller than the branching fractions of the electroweak vertices $B\to D\* D_s^{(*)}$, which are at the percent level.

We also note that, since $X(3872)$ can be produced via  a colour-favoured process, it would have a considerably larger production rate compared to $P_c$ states, if they are produced via colour-suppressed weak decay. Yet the experimental upper limit on $\mathcal B(B^+\to X(3872)K^+)$ is comparable to the lower limits on  $\mathcal B (\Lambda_b^0\to P_c^+ K^-)$ discussed above. (The lifetimes of $B$ and $\Lambda_b$ are comparable, so comparing branching fractions is similar to comparing decay widths.)

The situation worsens if we consider estimates for $\mathcal B (P_c^+\to J/\psi\, p)$ from models (Sec.~\ref{sec:rearr}). For a $3/2^-$  $\S\D^*$ molecule, Eides~\etal~\cite{Eides:2018lqg} predict
\begin{align}
\Gamma(P_c^+\to J/\psi\, p)=30~\textrm{keV},
\end{align}
and taking $P_c(4440)$ as an example, this is equivalent to
\begin{align}
\mathcal B (P_c^+\to J/\psi\, p)=1.5\cdot 10^{-3}.
\end{align}
From Fig.~\ref{fig:bfs}, this implies $\mathcal B (\Lambda_b^0\to P_c^+ K^-)$ is more than $10^{-3}$, strikingly inconsistent with  the weak vertex ~(\ref{eq:cs}) and the previous analogy with $X(3872)$. With our own, smaller prediction for $\Gamma(P_c^+\to J/\psi\, p)$,
the inconsistency is even more dramatic, implying  $\mathcal B (\Lambda_b^0\to P_c^+ K^-)\approx 7\%$.

As noted above, coupled-channel dynamics may resolve the discrepancy in the production of $P_c$ states, since they may be produced through components, such as  $\Lc\D\*$, accessible to colour-favoured weak decays. Given the scale of the electroweak vertices (\ref{eq:cf1}) and (\ref{eq:cf2}), it is plausible that this mechanism could enhance $\mathcal B (\Lambda_b^0\to P_c^+ K^-)$ to the $\mathcal O(10^{-4})$ level required by the GlueX result for the three new states. The $\mathcal O(10^{-3}- 10^{-2})$ values implied by models for $\mathcal B (P_c^+\to J/\psi\, p)$ are more problematic, so experimental measurement of the latter branching fraction would be helpful in determining the scale of the problem.

In our current model for $P_c(4440/4457)$, the only colour-favoured channel is $\Ll\D$. The predicted \cite{Chua:2019yqh} branching fractions 
\begin{align}
{\cal B}(\Lambda_b\to \Lc(2595) D_s^-)&=1.8\cdot 10^{-3},\\
{\cal B}(\Lambda_b\to \Lc(2595) D_s^{*-})&=2.5\cdot 10^{-3},
\end{align}
while still considerably larger than the colour-suppressed transitions, are smaller than the dominant colour-favoured transitions involving $\Lc\D\*$. Dealing properly with the production problems therefore requires a more thorough coupled-channel calculation including these dominant $\Lc\D\*$ states. 

Nevertheless, our current model is qualitatively consistent with data, in the sense that it allows for the comparable experimental fit fractions (\ref{eq:rats}) for $P_c(4440)$ and $P_c(4457)$. This is because we expect
\begin{equation}
{\mathcal B}(\Lambda_b\to P_c(4457)^+ K^- )  > {\mathcal B}(\Lambda_b \to P_c(4440)^+ K^-),
\end{equation}
due to the enhanced production of the $\LL\D$ component compared to $\S\D^*$, whereas
\begin{equation}
{\mathcal B}(P_c(4457)^+ \to J/\psi\, p ) <  {\mathcal B}(P_c(4440)^+ \to J/\psi\, p ),
\end{equation} assuming the inequality (\ref{eq:of}) in the partial widths is strong enough to translate, given the relative total widths, into an inequality in the branching fractions.
Presumably the two effects cancel each other to within a factor of two, yielding the experimental ratio of ratios of Eqn. (\ref{eq:rats}).

%

\section{Conclusions}\label{sec:conclusion}

We propose a molecular model of the new LHCb states, $P_c(4557)$ and $P_c(4440)$, that is  based on  long range pion exchange interactions, including the novel vector potential coupling the nearly degenerate $\S\D^*$ and $\Ll\D$ channels. Inclusion of the latter channel is crucial in generating two states in the appropriate mass region, and leads to the striking prediction that $P_c(4457)$ has quantum numbers $J^P = 1/2^+$. In our model, $P_c(4440)$ has $J^P= 3/2^-$, and no binding is predicted in the $J^P=1/2^-$ channel. Measuring these quantum numbers will serve to separate many of the proposed models. In particular, the $1/2^+$ assignment for $P_c(4457)$ is unique among models, which almost exclusively assign the state to $1/2^-$ or $3/2^-$.

Our model also gives a natural explanation for the relative widths of the states. The heavier state $P_c(4457)$ is narrower because its dominant $\Ll\D$ component does not couple via one-pion exchange to any lower-lying open charm pairs. On the other hand, the $P_c(4440)$ is dominated by $\S\D^*$ and thus couples strongly to a number of lower-lying open-charm channels, and we find that such decays dominate the total width. Among three-body modes, we predict that $P_c(4457)$  decays to $\S\D^0\pi$, whereas $P_c(4440)$ decays to $\Lc\D^*\pi$; experimental observation of these decay modes would be a  useful test of our model.

As for the total widths, we predict that $P_c(4457)$ should have a width of order 10~MeV, while the $P_c(4440)$ should have a width of approximately 100~MeV. The latter is too large. It is likely that this is due to the relatively large value of the combined coupling, $\hat g= 0.52$, which we have obtained in fitting the data. Indeed, the value of $\hat g$ that is preferred by baryon decays is about a factor of three smaller, which brings the predicted $P_c(4440)$ width near the experimental value. Our model has neglected short range dynamics entirely (as these cannot be determined at present); it is likely that these exist, and if they are attractive, that a quantitatively accurate description of the new states can be obtained. We explore this possibility in future work, where we also consider the larger system of coupled channels including $\LL\D$ along with all $\S\*\D\*$ and $\Lc\D\*$ channels, and $\Lc(2625)\D$~\cite{bs}.

We have also shown that the combination of LHCb and GlueX data provides a challenge for most models of $P_c$ states. The $\S\*\D\*$ components advocated in most models can only be produced in $\Lambda_b$ decays through either isospin-breaking or colour-suppressed weak decays. This implies considerable tension with existing experimental data. Coupled-channel effects can resolve this issue, as the states can then be produced through components, such as $\Lc\D\*$ or $\Ll\D\*$, accessible through colour-favoured decays.  Because of this we are able to find plausible explanations for the production characteristics of the putative pentaquark states near 4450 MeV, unlike other models. In our model, the production and $J/\psi\, p$ decays of $P_c(4457)$ are respectively enhanced and suppressed compared to those of $P_c(4440)$: this is qualitatively in agreement with the LHCb fit fractions $\mathcal R$, which are comparable for the two states.

Within our model charge splitting for the $P_c(4457)$ will be driven by the $D$ meson mass differences that exist due to the $\LL\bar D$ channel. Thus we predict that $P_c(1/2^+)^0$ has a mass of 4462 MeV. Note that the relatively large splitting would be very difficult to accommodate in compact pentaquark models. 

We also predict that isospin mixing will be minimal in both states, which also distinguishes this model from many others. 

Finally, we would like to emphasise that our model does not appear to suffer from the problem, which often arises in models of exotic hadrons, of predicting a proliferation of states which are not seen in experiment.  In the \icc~system we find only two states; in particular, there is no $J^P=1/2^-$ state, nor any $I=3/2$ states for any $J^P$. But we also need to consider whether additional states may arise once we include coupling to channels other than $\S\D^*$ and $\LL\D$, such as $\S\*\D\*$, $\Lc\D\*$ and $\Lc(2625)\D$. We discuss this in more detail elsewhere \cite{bs}, and restrict ourselves here to some general observations.

States with $I=3/2$ are not expected. This is because in $I=3/2$, potentials coupling $\S\*\D\*$ channels are weaker (suppressed by a factor of $-1/2$) compared to the corresponding $I=1/2$ systems, and also because the $\Lc\D\*$, $\LL\D\*$ and $\Lc(2625)\D\*$ channels do not contribute. (By contrast, in the simplest pentaquark models, $I=1/2$ states would be accompanied by $I=3/2$ partners.)

For $I=1/2$, we comment first on the negative parity sector, in which the dominant channels are those which couple in S-wave, namely $\S\*\D\*$ and $\Lc\D\*$. We expect a $J^P=5/2^-$ state below $\S^*\D^*$ threshold, as found in several other papers. It emerges naturally in pion-exchange models, as its elastic central potential in S-wave is more attractive than any other $\S\*\D^*$ channels. As remarked in Section~\ref{sec:elastic}, the apparent absence of this state in experiment has a natural explanation.

Whereas the $5/2^-$ system has only one S-wave channel ($\S^*\D^*$), the $1/2^-$ and $3/2^-$ systems have several S-wave $\Lc\D\*$ and $\S\*\D\*$ channels, so coupled-channel effects are more important, and strong conclusions are not possible without a detailed calculation. Nevertheless we note that among the $1/2^-$ and $3/2^-$ elastic channels, it is only $\S\D^*~(3/2^-)$, namely the channel which dominates our $P_c(4440)$, which has an  attractive S-wave central potential. The other channels $\S\D^*~(1/2^-)$, $\S^*\D^*~(1/2^-)$ and $\S^*\D^*~(3/2^-)$ are all repulsive, and are therefore less likely to support molecular states. Insight from the binding calculation without coupled-channels (Sec.~\ref{sec:elastic}) indicates that channels with repulsive potentials can support molecular states, but only with an unnaturally large value for the cut-off, which also renders other states too deeply bound in comparison to experiment. In this sense, consistency with experimental data implies that the molecular spectrum is restricted. (Again, this is quite different from the simplest compact pentaquark scenarios, in which the existence of one state naturally implies the existence of many others.)

The main novelty in our paper is the positive parity sector, where a $1/2^+$ state arises from $\LL\D$ in S-wave, coupled to $\S\D^*$ via the vector potential. It is natural to consider whether there are also analogous states formed from other combinations of hadrons. The closest analogue would be a $3/2^+$ state, with $\Lc(2625)\D$ in S-wave coupled to $\S^*\D^*$ via the vector potential. But unlike the \icc~ system, where the thresholds are almost exactly degenerate, in the $\Lc(2625)\D-\S^*\D^*$ system there is a considerable mass gap separating the thresholds, and on general grounds this implies a weaker potential. Considering that our $1/2^+$ state is scarcely bound, we therefore do not expect binding in the $3/2^+$ analogue. Similarly, we do not expect states with $\Lc(2595)\D^*$ or $\Lc(2625)\D^*$ constituents, as in these cases the nearest thresholds coupled by the vector potential, respectively $\S\D^*$ and $\S^*\D^*$, are separated by an even larger energy gap.

Indeed, the proximity of the thresholds of the channels \icc~coupled by the vector potential appears to be quite unique. In particular it does not hold in other flavour sectors, so we do not expect direct analogues of $P_c(4440)$ or $P_c(4457)$ with hidden strange or bottom. Another unique feature of our \icc~ system, which is important in the formation of bound states, is that all four of the interacting hadrons are narrow: this is not generally true of systems with a vector potential, which necessarily involve at least one P-wave hadron.

Geng~\etal\cite{Geng:2017hxc} have searched more systematically for other systems with a vector potential which could potentially support bound states, considering in particular the requirements that the thresholds are approximately degenerate, and that the constituent hadrons are narrow. The only other candidate they find is the 	$\Lc(2595)\bar\Xi_b-\S\bar\Xi_b'$ system. Aside from this possibility, which seems experimentally challenging, it appears that the \icc~system considered in this paper is quite unique.

\acknowledgments

The authors are grateful to Jun He, Tomasz Skwarnicki, Yuki Shimizu, Lubomir Pentchev, and Manuel Pavon Valderrama for discussions.
Swanson's research was supported by the U.S. Department of Energy under contract DE-SC0019232.

\end{document}